# Constructing Strategy of Online Learning in Higher Education: Transaction Cost Economy

Yabit Alas [#1] and Muhammad Anshari [#2]

Universiti Brunei Darussalam
*Brunei Darussalam*

[1]yabit.alas@ubd.edu.bn
[2]anshari.ali@ubd.edu.bn

**Abstract**

*The online learning tools and management also known as Learning Management System (LMS) have been adopted by higher education as it allows convenient and flexibility in learning process between students and instructors or tutors with minimal cost. The adoption of online learning tools in university has allowed users (students and instructors) to interact, share and discuss anytime-anywhere conveniently. Many students nowadays rely on online resources based using their mobile devices, substituting traditional learning interactions. Universities need strategy to sustain in providing intensive interactions and spreading word out mouth of good services through online learning tools by focusing on niche markets and creating close relationship with their stakeholders. The study presented in this paper analyses how universities design best practices in adopting LMS and evaluate its current state for future improvement. In fact, with proper strategies of LMS, universities have opportunities to sustain their business by offering interesting packages and to improve their services through intensive interactions with their users. In this study, we deploy Transaction Cost Economics (TCE) to understand the change business environment and to construct a model for higher institution to regulate their scenario on online learning strategies in fast changing and threatening business environment.*

## 1. Introduction

Communities and stakeholders in higher education nowadays have gained more bargaining power in term of information as they can have access to multiple online resources (Almunawar et al, 2013a). In a high competition landscape, customers can demand to educational providers of online learning tools that are aligned with their preferences, life style, and recent technological trends. For instance, students and instructors hesitate to use online learning tool, which is not user friendly though the tool is free for use and considerably customizable. Thus, providing friendly learning management service (LMS) yet cost efficient and effective without reducing core functionalities are unavoidable for any education providers. In fact, the trend shows many universities have adopted e-learning tools since last decades with different level of complexities (Low & Anshari, 2013). Universities deploy any form of online learning tools because it promises benefits such as breaks the space and time for travelling to stay comprehensive and competitive in its core business (learning, teaching and research).

The advancement of any online learning tools have significantly affected by the changing customers behaviour in absorbing information from multi layers interactions driven by the advancement of ICT (Anshari et al, 2013a). With the steady tendency of using mobile technology, social networks and Web 2.0, education providers can improve their services. Deploying social networks into the learning process shows empirical evidence that blogs and wikis have the potential to support internal communication (Calvó-Armengol et al, 2009) and knowledge sharing (Chow & Chan, 2008). Web 2.0 can sustain the process of interactive learning (Baxter, Connolly and Stansfield, 2010; Boateng, Mbarika and Thomas, 2010) where organization may extract information from conversation that takes place in the social networks.

There are three main actors in delivering LMS; university management as service provider, students and instructors as users, and system provider as developers or operators either third party or internal IT team. Management plans the implementation, decides system that is in line with strategic goals, evaluate the progress, and revises for improvement. System providers have various functions such as development, customization, and maintenance. While, students and instructors use the system and provide feedback for improvement (Almunawar et al, 2013b). All these functions are heavily depended on information technology and strategic planning and implementation. Choosing appropriate system for online learning tools can be daunting task for universities. On the other hand, there are growing number users with high information technology and Internet literacy. With rapid growth of mobile technology, users as well as IT savvy generations the adoption of Web and Apps based learning services are very high (Anshari et al, 2013b). More and more people adopting online learning systems as these system offer efficiency and convenience as compared to the traditional learning. Therefore, both service provider and users have common interest that the service should be faster, cheaper, reliable, and secure. Transformation of learning process from physical based learning interactions to blended learning (online and physical) interactions will likely to happen.

This demand eventually will force system provider to have a proper system and well as integrated strategy to foster sustainability (Prencipe et al, 2003). If there is no proper business' strategy to face the serious threat from technology side that is always changing, system providers have slim change of survive. The growing number of system providers offering similar services is now forcing the vendors to be more responsive and competitive to survive. Either commercial or open source system providers need to sustain their business in providing learning management system (LMS) must seek to create long-term value by embracing the opportunities and risks related to the protection, enhancement, and sustainment of the important resources.

Our study is aimed to figure out the recent trend of online learning tools or LMS adoption in higher education level. The Transaction Cost Economics (TCE) is usedto analyzethe changing of business environment. TCE provides the guidelines to determine which structure would be appropriate for which transactions required for the performance of the task. Based on the trend and theory mapping, we propose a model for alternative online learning solution in higher education. In the next section, we present a literature review of related work, and Section 3 contains the methodology of our research. We present our discussion in Section 4. Finally, Section 5 presents the conclusion.

## 2. Literature Analysis

*ICT and Education*

Over the past few years, the use of ICT has integrated into education to reach students. Students who always connected to the Internet through their smart mobile devices in the classroom will be facilitated with the integration of ICT in education. On the other hand, the institution has pushed to integrate latest technologies into their study mode, contents delivery, management, and curriculum. The adoption can benefit institution to strengthen its position in community and at the same time as marketing strategy for institution in advanced ICT.

Briefly, the timeline portrays from the early of the computer was used in education when MIT used computer for a flight simulator to trained pilots.Then IBM 650 released the first commercialized computer in the market. In 1967, Apple computer introduced Apple II also known the personal computer.

In 1996, the evolution continued one in every twelve students in US had computer access (Arora, 2013). The Internet boom in 1990s was another milestone for asynchronous forms of education through distance learning has drawn the attention of students. Asynchronous is a mode of content delivery where participants are not required to access materials at the same time but on their own schedule. For example, asynchronous can be in the forms of email, voice mail, audio or video recording. On the other hand, synchronous technology requires participants to be present at the same time such as video conferencing, Internet radio, live streaming, etc. In 2007, one in every five-university students was enrolling online education either synchronous or asynchronous mode (Arora, 2013). The other education technology is adaptive learning. Online education system modifies the presentation of each student contents in response to observes aspects of student performance (Tseng et al, 2008).

As students become more attached to ICT, variety of tools that leveraged users to interact, generate contents, and share information in social media platforms. It is believed that the adoption Web 2.0 or social network in education setting can increase in students' participation. For instance, less sociable students have become more confident and participatory to ensure all students engaged. However, there is less imperative evidence that there is correlation between social network and students' performance. Furthermore, adoption of mobile learning (mLearning) in education gives opportunity for students to learn anywhere and anytime beyond the classroom.

The other challenges is classroom becomes more technological overwhelmed where students who always carry and depend on a computer or a smart mobile device that is used on a regular basis. ICT offer students to browse for information quickly and allow them to collaborate on projects in their learning process, offer apps schoolwork. However, they should consider seeing technology as a means to enhance learning not temporal short-term ideas. In addition, social networks allow students to maintain interaction but it is lack emotional relation compare to face-to-face interaction. Then, how do institutions draw a line to ensure technological demands, social balance, and educational achievement.

Finally, in term of adoption, the other trend in educational technology is the cloud-computing mode. Cloud computing in LMS has fundamentally changed the higher institution on adopting online learning tool. Cloud computing has ability to adopt the latest online learning system without investing in IT infrastructure. It offers reliability of service because resources managed in a trusted environment.

### *ICT and Institution*

How ICT is transforming higher education? Universities show utilizing advance ICT to revolutionize the way to deliver the knowledge and contents. For example, online degree program is a niche channel of expending revenue opportunities to enable those who cannot attend a degree through brick and mortal mode. The two type of distance learning, synchronous and asynchronous, address flexibility of learning model that often implemented in parallel to traditional face-to-face mode.

In addition, Web 2.0 provides opportunity for researchers or research groups in universities for collaboration and knowledge sharing. It helps to find collaborators for research and possibility to communicate with other researchers in the same research cluster. *Researchgate* is one of the scholarly website that accommodates Web 2.0 technologies that facilitates interactivity, comments on articles, and open peer review process among scholars.

Regardless of the benefits, technology is still a disruptive innovation and expensive investment for some higher institutions (Jayson, 2013). Many higher institutions deploy "technology in trend" is not merely for their business or functionalities, but institutions require to demonstrate a commitment towards advanced ICT in order to attract community trust, better branding and image, partnership and corporate funding.

**Transaction Cost Economy (TCE)**

The TCE characterizes as set of activities connected by transactions. The activity is the part of a service, while a transaction is a series of the activity when one activity ends and another one begins. Therefore, the relation between activities occurs when services are shifted between points to another. In addition, the transactions costs are affected by a market based system and correlated with the allocation of tasks to external factors (Bello, Dant & Lohtia, 1997).

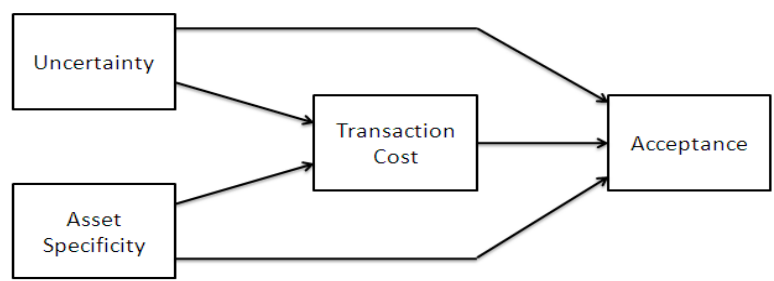

Figure 1: TCE Model (Liang and Huang, 1998)

In figure 1, Liang and Huang (1998) propose the transaction cost that could be determined by asset specificity and uncertainty in order to achieve users' acceptance. Asset specificity is mainly in a business scenario like buyer-seller situation, where the buyer is the party that does not hold the specific assets and the seller is the party that holds the specific assets. In the case of university, the asset specificity is the lecturers, experts, tutors as factory of brains to deliver ranges of knowledge. Asset specificity can be the key to organizational survival is the ability to acquire and maintain resources. Joskow (1988) argues the community, corporate, government agencies (the party that does not hold the specific assets) have exit cost associated with time and searching investment if they decide to switch party.

For instance, students may consider online learning program if the transaction cost is lower than face-to-face learning. The cost can incur for time and searching the independent study for online learning program. Uncertainty is any factors that affect to the transaction cost. TCE discusses on how the external uncertainty as one of critical resource to the university. The fast changing technology in learning environment can create uncertainty for the university. Users preference may change over time that create uncertainty for the organization. Therefore, the university needs to find niche market to manage external uncertainty.

### 3. Methodology

By looking into current students' adoption of technology and demands for reliable online learning tool, the study attempts to propose online learning strategy in higher education scenarios. For this purpose, we designed survey to observe the expectation of users. We use the purposive sampling methods in which participants were selected from first semester university students who is new in using online learning tool in university. Perception of online learning tools were analysed in order to understand the expectation. Data gathered from online survey was examined, interpreted and eventually converted in requirements to develop a strategy for higher education. The model will be further examined to understand the differences and for future recommendations. The following are the profiles of our respondents. Table 1 shows the demographic characteristics of the participants. Majority of the participants are local Bruneians, who study at the national university in the country. The participants range from sixteen to thirty years old. Therefore, they represent young generation in the country. The time taken to complete the questionnaire on average was 5 minutes. There were 91 students participating in the survey, which was conducted from July to August 2014. While, 57% of them access Internet more than 4 hours per day, and 23% of participants' access 2 to 3 hours daily, and remaining is less than 2 hours.

| **Item** |  | **Percentage** |
|---|---|---|
| **Gender** | Male | 21% |
|  | Female | 79% |
| **Age** | 16 - 20 | 91% |
|  | 21 - 30 | 9% |
| **Internet Usage** | Less than 1 hour daily | 6% |
|  | 1 – 2 hours daily | 14% |
|  | 2 – 3 hours daily | 23% |
|  | More than 4 hours daily | 57% |
| **District** | Brunei - Muara | 73% |
|  | Tutong | 12% |
|  | Belait | 13% |

| | Temburong | 2% |
|---|---|---|

Table 1: Demographic of Participants

## 4. Results and Findings

In this section, we discuss the findings in regards to the participants' opinion on online learning and study's performance. When we ask about online learning would make them easier to learn show that 44% of them agree, 51% is not sure yet, and 5% of participants disagree. It indicates that more than half of participants cannot see the benefit of online learning tool in relation to support their study. Similarly, when we ask a correlation between online learning and student's achievement, the result is 22% of participants agree with the statements whereas majority of participants are not sure either online learning can make them obtain a better grade.

Next, we asked them about the nature of online learning enables them to study any time. Majority of participants (72%) agree while the other 25% abstained. Respondents who prefer online learning believe that online learning systems are convenient and time saving. Online learning is convenient as learning can be done anywhere and anytime through clicks of mouse or touch of fingers. In addition, online learning systems are very efficient as learning through the Web or Apps only requires students to access on the website. It confirms with the majority of participants, where they are young, and highly Internet connected users that they like to use online media but they are not sure about effect of online learning system and their study performance. We ask further their opinion on the urgency of face-to-face interaction with instructor. 80% of respondents agree that they need face-to-face interaction. We also interested to ask about the use of online learning tool to share info and discuss subject matters where 60% of participants would use the facilities to share and discuss about the subjects.

| Item | Agree (%) | Not sure (%) | Disagree (%) |
|---|---|---|---|
| Online learning will make it easier for me to learn | 44 | 51 | 5 |
| Online learning will help me to obtain a better grade | 22 | 73 | 5 |
| I like online learning because it is not limited to regular school hours | 72 | 25 | 3 |
| Face-to-face interaction with my instructors and friends is very important | 80 | 19 | 1 |

| I like to use online learning tool to share info and discuss subject' matters | 60 | 35 | 5 |

Table: Survey Results

Our survey indicates that there is slow trend of shifting students' habit from the traditional of face-to-face learning to the online learning mode. Figure 2 portrays the scenario of the shifting habit for students in adopting online learning systems. The survey revealed the students who enroll for full time at university prefers to have brick and mortar system because they are comfortable to meet and discuss with the instructors directly and they have not engaged in any full time or part time jobs other than going to university. The availability of online learning system is perceived as complementary to support their studies. On the other hand, the trends prevail for online learning system for a niche market of students who have secured their work without leaving the job yet they can enroll for online courses or students who live far in distances from university are benefited from the service.

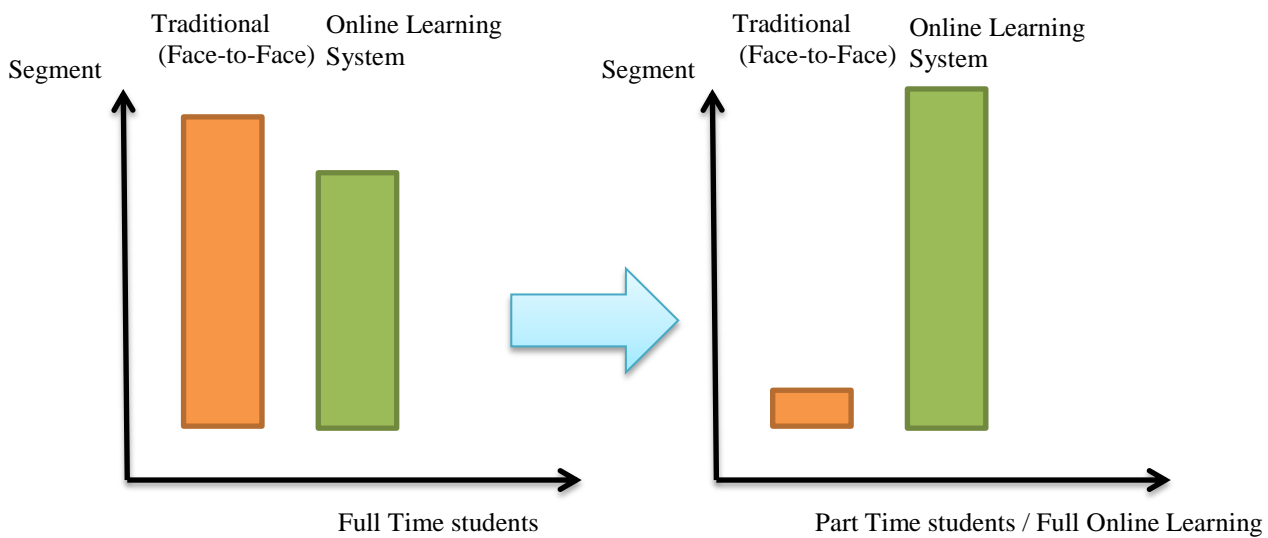

Figure 2: Scenario of students' habit on online learning system

The results send an alarming message to the higher education that they are facing opportunities on the students' preference and their service quality. To help the extendibility of knowledge sharing to the community we propose that they consider the innovation of online learning system toward service excellence.

Online learning systems are very popular and convenience for some students who are inflexible with their time arrangement. However, blended approach is a possible alternative for higher education in providing learning solution. The reasons are the advantages of blended personal touch and personal tech as supportive tools. Personal touch means that students are able to directly ask any queries, get immediate responses, and accept the advice from instructors to avoid miscommunication and more conformity. In addition, traditional learning systems offer real full ranges of students' life that promise rich and real experiences.

5. Discussion

Although the adoption of LMS is increasing worldwide, many students are still in need of the face-to-face interaction, because they offer some distinctive benefits such as less barrier in knowledge transfer, conformity, and conducive learning environment. In addition, there are several perceived disadvantages of online learning that affect students' confidence including security and privacy issues, information overload, distraction, and lack of personal touch or human interactions. With the background of the students who are mostly in the high level of Internet literacy, then the students are categorised into their perception between those who trust and confident with online learning system and students who uncomfortable using online learning system. Both groups have different modes in terms of arranging their learning experience. Internet savvy students use the web or Apps more heavily to arrange their learning. They believe that the Internet offers them more options and the ability to enrich knowledge through collaborative sharing. However, when learning becomes more complex and involves extensive consultations, they normally seek out face-to-face consultation. As for students with a lack of trust and confident with online learning system because they perceive many distraction and less focus compare to attend to the class physically, they hesitate to use the Web or Apps in learning process only for basic functionalities like retrieving online materials from the portal.

There is no single solution fits for all scenarios, higher institution should carefully decide at which direction that the institution is going to upgrade for online learning system before understanding the needs of users towards the systems. From the survey revealed some interesting keywords to explore. First, they are youth with moderate to advance level of computer literacy and highly connected to Internet. They mostly access Internet from both personal computer and smart mobile. Secondly, though they are Internet savvy but majority of the students yet prefer face-to-face interaction. They need LMS as supporting tools of their learning experience at university and complementary for their learning process. Third, the management views that implementing or upgrading LMS is required to fully utilizing the latest educational technology so that the students are experienced and exposed to boost the learning outcomes. In addition, the reason for deployment can be derived from internal marketing strategy to acquire niche market on online learning segment. Universities deploy the LMS to extend additional services to automate some of their core operations in teaching, learning and research so that they can increase profits through online learning value added services. LMS enables students from passive information receivers to active participants in information creation, learning process, and collaboration. In addition, efficiency and convenience offered by LMS may pose serious threats to traditional learning especially for those that do not have online strategies to tackle the threats. It is interesting to find out how university respond to the changing business environment that challenges their sustainability.

**Transaction Cost**

TCE is used to analyse the implementation of online learning system because the theory offers understanding in managing uncertainties and establish asset specificity. Managing uncertainties inferred that university is determined largely by the external factors. Since it is determined by the external factors, university needs to perform strategies that allow them to acquire these resources. While, asset specificity is uniqueness of the assets or high specificity of assets will affect transaction costs higher. In case of university, the asset specificity can be knowledge expertise and availability of extended value system (LMS). For instance, specific asset like knowledge expertise in universities is in a vulnerable position. Potential students can be forced to pay higher cost if there is a limited learning service and students can pay a lower transaction cost if there are many options to acquire learning like online learning service.

TCE proposes to manage uncertainties is to develop a specialty or offer exceptional service to those uncomfortable with traditional learning process. In that sense, university increases its likelihood by selecting more controllable niches in which to do business. For example, online learning system can secure business on corporate training packages of the government and the private sector, which regularly assign their officers to upgrade their knowledge and learning specific courses. In the case of online learning system can take place whole year around indeed is a potential market. Students can make own arrangement to start online learning. Instead, they choose their interested courses to enrol and pay the package in cash. In addition, transaction cost can be reduced through establishing external linkages to manipulate exchange relationships (Pfeffer & Salancik, 1978). In this scenario, the cloud computing can be the best option to deliver the online service quickly into the market with cost efficiency.

It is also acknowledged that in overall of the transaction cost for online learning can be lower especially for those who really have dedication and commitment with the online learning program' structure. The availability of a Web and Apps anytime-anywhere is the most appealing sources of information for online students. For students who make personal arrangement in learning may bear search transaction cost in finding the best course that suit their need but the cost can be compensated with the accelerate learning process compare with those traditional learning that is fixed in time and process.

**The Model**

Based on the literature review and survey results, we propose a model operates in the domains of uncertainty and asset specificity to achieve students' acceptance. Figure 3 depicts the transaction cost model on online learning system. Firstly, university need to list their asset specificity, expertise in the university is asset specificity that must be managed properly. They are "factory of brains" in online learning program than can deliver teaching and learning program without the borders. Though, the most challenging part in managing knowledge expertise are to transform culturally to utilize online learning tool for knowledge transfer ability.

Secondly, manage uncertainty through securing niche market. Uncertainties can come from advancement of technology or changing in users' preferences. In order to manage uncertainty, university maintain online learning distinct market or unique market segmentations. Market segmentations rely on creativity in promoting and convincing their online learning service. University must maintain their niche markets while at the same time improving their core traditional learning to maintain students' loyalty.

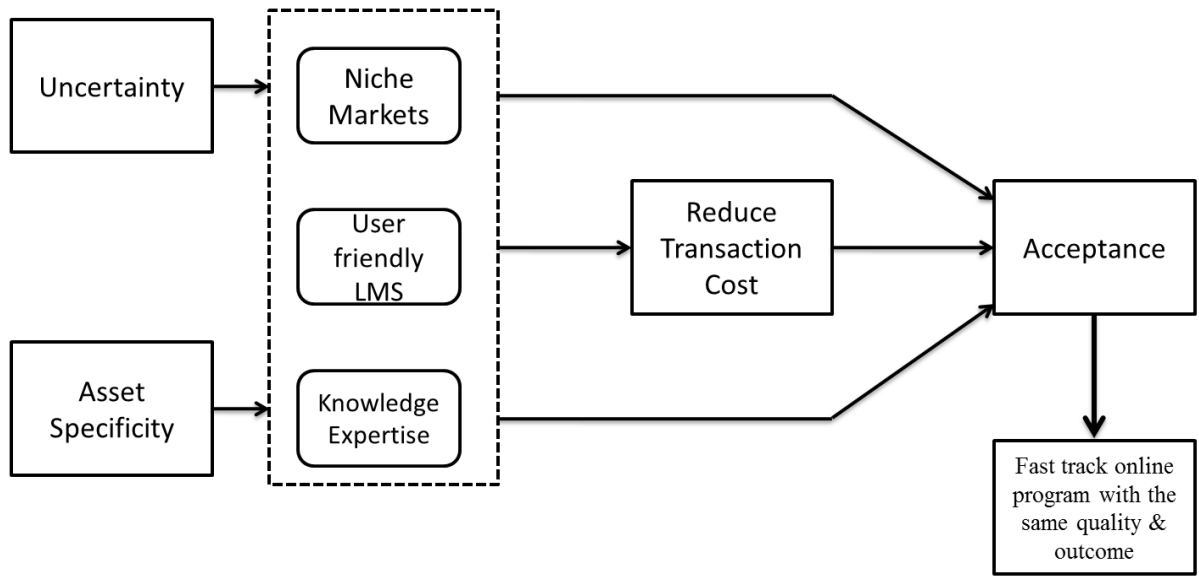

Figure 3: TCE for Online Learning Program

Thirdly, accommodating LMS that is user friendly plays a critical role in creating asset specificity, which escalates business operations. Indeed, the LMS with Apps enabled are powerful channels for university to develop, enhance interactions and implement relationship practices with students. It facilitates peer-to-peer collaboration and easy access to real time communication (Almunawar & Anshari, 2011). Finally, TCE proposes that university may consider online learning program when they can secure the factor of uncertainty and manage asset specificity that will reduce total transaction cost for all parties. However, the acceptance of the program will likely to improve if the quality and outcome of the students who enrol in online learning program are at least similar or even better compared with the traditional face-to face learning. Online learning program offers fast tract learning program that can accelerate overall learning process, which is in turn, reduce overall transaction cost. To sum it all briefly, recent LMS has allowed virtual interactions, providing tools that emulate human skills and knowledge to cater to student preferences and to match an individual's requirement. The virtual model of learning is flexible as to accommodate the changing learning environments.

## 6. Conclusion

The adoption of online learning program in education has been promising a new experience to the students, education provider, instructors, and community. In fact, learning technological innovations will continue to have major effects on teaching approaches over the coming years. University has successfully secured their markets through mastering core learning service. However, there is a very high tendency of individual who would prefer to use online learning system for their learning process. Therefore, the potential loss of potential students who prefer online learning, if not taken care of properly, is very high. It is important for university to sense the future impact of online learning systems and plan appropriate online strategies for extending their learning service to broader community. On the other hand, university should also realize that online learning system could be beneficial as well as unsuccessful. It creates opportunities for collaborative learning and extends the market for university but at the same time, it is also a big challenge that needs proper strategies to neutralize when dealing with the social aspect of learners.